\documentclass[3p,twocolumn]{elsarticle}
\usepackage{graphicx}
\usepackage{dcolumn}
\usepackage{bm}
\usepackage{natbib}
\usepackage{color}
\usepackage{float}
\usepackage{amsmath}
\usepackage[T1]{fontenc}
\def \gmt {Ge$_{1-x}$Mn$_x$Te}
\def \smt {Sn$_{1-x}$Mn$_x$Te}
\def \pmt {Pb$_{1-x}$Mn$_x$Te}
\def \spmt {Sn$_{1-x-y}$Pb$_{y}$Mn$_x$Te}

\journal{Journal of Magnetizm and Magnetic Materials}

\begin{document}

\begin{frontmatter}

\title{Magnetic anisotropy energy in disordered \gmt}


\author[ifp]{A.~{\L}usakowski\corref{cor1}}
\ead{lusak@ifpan.edu.pl}

\author[ifp,byd]{P. Bogus{\l}awski}

\author[ifp]{T. Story}

\address[ifp]{Institute of Physics, Polish Academy of Sciences, Al. 
Lotnik\'{o}w 32/46, 02-668 Warsaw, Poland}

\address[byd]{Institute of Physics, University of Bydgoszcz, ul.
Chodkiewicza 30, 85-072 Bydgoszcz, Poland}

\cortext[cor1]{Corresponding author}

\begin{abstract}
We theoretically analyze the influence of chemical disorder on magnetic
anisotropy in \gmt\  semiconductor layers known to exhibit carrier-induced
ferromagnetism and ferroelectric distortion of rhombohedral crystal lattice.
Using DFT method we determine the local changes in the crystal structure due to
Mn ions substitution for Ge and due to the presence in \gmt\ of very high
concentration of cation vacancies. We calculate the effect of this structural
and chemical disorder on single ion magnetic  anisotropy mechanism and show that
its contribution is order of magnitude smaller as
compared to magnetic anisotropy mechanism originating from the spin polarization
induced by Mn ions into neighboring Te and Ge ions. 
We also discuss magnetic anisotropy effects due to pairs of Mn
ions differently allocated in the lattice. 
The spatial averaging over chemical disorder strongly reduces the strength of 
this magnetic anisotropy mechanism and restores the global rhombohedral symmetry 
of magnetic system. 
\end{abstract}


\end{frontmatter}


\section{Introduction}
\label{wstep}
GeTe belongs to the IV-VI family of narrow gap  semiconductors. At temperatures
below $T_0$ = 670~K its crystal structure has rhombohedral symmetry $C_{3v}$
which can be characterized by three parameters: the lattice constant
$a_0\approx 6$\AA, the rhombohedral angle $\alpha \approx 88.3^o$ and
the relative displacement of Te and Ge sublattices along [111]
crystallographic direction. This displacement which is equal to
$a_0\sqrt{3}\tau$, where $\tau\approx
0.03$, causes GeTe to be ferroelectric. \\
Manganese ions Mn$^{2+}$ introduce in \gmt\  local magnetic moments and this 
IV-VI
diluted magnetic (semimagnetic) semiconductor (DMS) exhibits ferromagnetic
transition with the 
Curie
temperature $T_C(x)$ up to 190~K \cite{lechner, fukuma}. Due to 
simultaneous ferroelectric and ferromagnetic ordering \gmt\  belongs to the 
class of multiferroic materials, which may become very interesting for
possible future applications.

Our previous theoretical analysis of the physical mechanisms of 
magnetic anisotropy in \gmt\ \cite{lusakowski0}, hereafter referred
as paper I,
was inspired by recent, unexpected 
results of magnetization and ferromagnetic resonance experiments
which clearly showed that the easy axis of magnetization in thin layers of \gmt
, for the manganese content of about 10 at. \%, is perpendicular to the
layer \cite{fukuma,knoff1,knoff2,fukuma1,przyby,knoff2015}. 
In ferromagnetic thin layers the shape anisotropy
usually dominates with easy magnetization axis located in the layer plane. This
is indeed observed in \gmt\ layers for higher content 
of manganese, above
 20~at. \%. For these layers the X-ray diffraction analysis (XRD) shows that 
the increase of 
Mn content results in transition from rhombohedral to cubic (rock-salt) crystal
structure.

In the paper I we performed extensive {\em ab initio} density functional 
theory (DFT) calculations of magnetic anisotropy energy (MAE) 
in \gmt. The aim
was to elucidate microscopic mechanisms  responsible for the magnetic 
anisotropy.
We concluded that the main contribution to MAE is given by the spin 
polarization 
induced in Te and Ge neighbors of Mn, and not by the Mn ions themselves. 
The Mn spin polarizes its neighborhood, the spins of Ge
ions are approximately parallel and the spins of Te ions approximately 
antiparallel to the direction of the Mn spin. 
The degree and the spatial range of spin polarization  
is relatively small in insulating crystal but it strongly grows  
with the increasing concentration of holes. 
Because of spin-orbit (S-O) interaction, the 
directions of Te and Ge spins forced by the direction of Mn spin do not,
in general, correspond to minimum of the total energy of the system. 
This forcing of Ge and Te polarizations by 
the Mn spin is the principal reason why variations of the directions of Mn 
spins result in changes of the total energy. 
Apart from the identification of the main physical mechanism responsible for
magnetic anisotropy we also showed that MAE is determined by the hole 
concentration, macroscopic crystal structure, and the local configuration of Mn.

The paper I was devoted almost exclusively to the situations where
there was  
only one Mn ion in $2\times2\times2$ rhombohedral supercell containing 64
atoms. 
In all cases, the angular dependence of MAE on the Mn spin direction 
characterized by the angle $\theta$ was well approximated by the formula 
\begin{equation}
 E_A(\theta)=a_{2c}\left(\cos 2\theta - 1\right).
 \label{eq1}
\end{equation} 
Here $\theta$ is the angle measured in the (-110) plane from the [111] 
direction. 
Thus, for $\theta=0$ and $\theta=90^{\circ}$, the Mn spin is parallel to [111] 
and to the [11-2] crystallographic directions, respectively. 
The same definition of $\theta$ is used in the present paper. 

One should keep in mind that a supercell containing one Mn ion simulates 
perfect, 
translationally invariant crystal with rhombohedral symmetry. 
In actual random substitutional alloys, this symmetry is locally destroyed. 

Although macroscopic structural, mechanical, electrical or magnetic properties 
can have 
the global rhombohedral symmetry as confirmed by XRD,  
at the unit cell scale this symmetry is destroyed. 
This is due to the randomness of the occupancies of lattice sites
in the chemically mixed cation sublattice (which is sometimes called chemical 
disorder), 
and to the presence of native defects, in our case the cation vacancies. 
The random positions of manganese ions together with
different ionic radii of Ge and Mn lead to local deformation of the
lattice. The local deformations are also caused by the germanium
vacancies.
Each cation vacancy, like in other IV-VI semiconductors, delivers two holes,
thus in real sample with the hole concentration of the order
$10^{21}$~cm$^{-3}$ we expect $5\times 10^{20}$~cm$^{-3}$
of vacancies.
For comparison, the Mn concentration $x=0.01$ corresponds to about
$1.9\times 10^{20}$ of Mn ions per cubic centimeter.

The influence of the chemical disorder on the magnetic anisotropy in \gmt\ is
the topic of the present paper. The paper is organized as follows. 

In Section \ref{sec2} we analyze the single ion magnetic anisotropy (SIMA) in
\gmt. In the paper I we explicitly showed that SIMA in the
conducting \gmt, is of minor importance compared to the anisotropy caused by
spin polarized free carriers. However, the calculations were performed for 
systems
with manganese ion's neighbourhood of perfect rhombohedral symmetry.
On the other hand it is well
known from numerous electron paramagnetic resonance experiments (EPR) that the 
ground
state splittings and consequently magnetic anisotropy properties of Mn$^{2+}$
ion strongly depend on the symmetry of its surrounding.  Previous
calculations performed for \pmt\ and \spmt\ show that the ground
state splitting of Mn$^{2+}$ ion in disordered environment may be of the order
of 1 Kelvin. Because the difference in ionic radii of Ge and Mn is much
smaller than that of Pb and Mn, the local lattice distortions (which strongly
influence the magnetic properties of Mn) are also smaller. Thus,
we do not expect that SIMA driven by the microscopic lattice distortions 
are dominant in \gmt. 
On the other hand, the band structure of GeTe is
significantly different from those of other IV-VI semiconductors due to
ferroelectric properties of the crystal. That is why direct calculations of SIMA
in \gmt\ are necessary. 

In Section \ref{sec2} we briefly describe mechanism responsible for SIMA and the
method 
of 
calculations. The obtained results confirm that even in the presence of 
disorder, 
in conducting \gmt\ SIMA is of secondary importance.

Section \ref{sec3} presents results of {\it ab initio} calculations of MAE 
of pairs of Mn ions in \gmt. All nonequivalent positions of two Mn in the 
$2\times 
2\times 2$ supercell are taken into account. We also show a few results for 
three Mn ions in the supercell. 
In the calculations the spins of Mn ions are always parallel. 
This is justified, because according to the results of 
Section II the energies connected with local magnetic anisotropy easy axes are 
small, and in the first approximation the possible noncollinearity of 
Mn spins can be neglected. 
We explicitly show that the direction of 
easy axis of magnetization strongly depends on the relative position of Mn 
ions, 
and MAE is not described by Eq.~(\ref{eq1}). For chosen
two configurations  of two manganese ions we analyze contributions of kinetic, 
electrostatic and
exchange-correlation energies to MAE. It is shown that the Mn ions
introduced to GeTe causes significant perturbation in spatial 
distribution of electron charge and, consequently, significant change
of electrostatic potential which, via the S-O interaction, influences MAE.

In Section \ref{sec4} we summarize the main conclusions of the paper. 

\section{Single ion magnetic anisotropy}
\label{sec2}
\subsection{Theory}
According to the Hund's rule, the ground state of Mn$^{2+}$ ion is 
$^6S$, with the orbital momentum $L=0$ and spin $S=5/2$.  As the
orbital singlet, it does not interact with the crystal environment, 
and its ground state without external magnetic field is six-fold degenerate. 
This model of Mn$^{2+}$ ion explains a number of 
phenomena, e.g., the results of magnetization measurements in 
diluted magnetic semiconductors. However, if we consider 
problems in which very small excitation energies are of importance, this model
is no longer valid. An example of such a 
problem is the splitting of the ground state of Mn ion in a crystal field  
measured in EPR experiments.
In the theoretical analysis for zero magnetic field, the Mn$^{2+}$ ion is 
described 
by a $6\times 6$ matrix, $H_{MM'}$, which plays a role of effective spin
Hamiltonian \cite{abragam}. 
The subscripts $-5/2 \le M,M'\le 5/2$ denote 
the projections of Mn spin on the quantization axis. For
example, for a perfect cubic symmetry of Mn surrounding
\begin{equation}
H = \frac{a}{6}(S_x^4+S_y^4+S_z^4).
\label{eq2}
\end{equation}
The lowest excitation energies of the ion (i.e., the eigenvalues of $H_{MM'}$) 
strongly depend on the symmetry of its 
neighborhood, and are in the range $10^{-4} - 10^{-1}$~K. 
If the Hamiltonian $H_{MM'}$ is not diagonal, it leads not only to the ground 
state
splitting but also to the magnetic anisotropy of the
ion \cite{abragam,yosida}.
\\
Density functional theory, which is one of the basic theoretical
methods used to study condensed matter properties, is not suitable for this 
problem
because DFT calculations determine the electron density in the ground state, and
provide no direct information about the excited states of the system,
particularly in situations where the energies of the excited states are very
small. In this case one should use methods based on perturbation theory. 

There are two mechanisms known in the literature which lead
to the ground state splitting: interaction with excited states of $3d^5$
configuration, and
hybridization between $3d$ and orbitals of the surrounding ions. 
In the first mechanism \cite{sharma} the internal manganese S-O interaction 
couples
$^6S$ with the excited $^4P$ states, due to which the ground
state of Mn is no longer a pure $S$ state with total angular momentum $L=0$.
Consequently, the interaction with the crystal environment becomes possible.  
As a result, ground state splitting and nondiagonal effective Hamiltonian 
$H_{MM'}$ are finite.

In DMS, the second mechanism, 
i.e.,  the hybridization between $3d$(Mn) orbitals and the orbitals of the 
neighboring ions, plays a more important role. 
In this case, a system consisting of $3d$(Mn) electrons and valence  
band carriers is considered. 
In the ground state of this system, which is six-fold degenerate due to 
manganese spin $S=5/2$, there are five 
electrons on the $3d$ shell and the host bands are filled up to the Fermi 
energy.  
In the excited state, one
electron is transferred from the host band to the $3d$ shell (or 
from the $3d$ shell to the band) resulting in one additional hole 
(or electron) in the host bands, respectively. 
Such virtual transfers of electron
were used earlier in theories of $sp-d$ exchange interaction in
semiconductors \cite{larson,sliwa,kacman}. As a consequence of this $3d$ shell -
band states 
 hybridization, the ground state of the system is no longer six 
fold degenerate, and the resulting low energy excitation spectrum and the 
anisotropic properties of the ion take place, in accord with experimental  
observations. 
This mechanism
was described in Ref. \cite{lusakowski2} and successfully applied to
EPR results for PbTe:Mn layers grown on BaF$_2$ and KCl
substrates \cite{lusakowski2}, to the single ion anisotropy in \spmt\  mixed
crystals \cite{lusak_ssc} and to the analysis of magnetic specific heat in
\pmt\ \cite{lusakowski3}. 
This method is also used in the present paper for calculations of effective
spin Hamiltonian. 
Here, we do not repeat detailed description
of the method and formulas contained in Ref.~\cite{lusakowski2} but
sketch only the main idea of calculations.

We consider GeTe crystal at the temperature $T=0$ with one germanium atom
replaced by manganese atom. It is assumed that this replacement does not change
appreciably neither the band structure nor the spatial electron distribution.
The only difference comparing to the pure GeTe is the presence of additional
$3d$ shell on one of the cation places with five electrons on it. If we neglect
hybridization between $3d$ 
orbitals and the orbitals of the neighbouring atoms, in the ground state of such 
a
system there are five electrons on $3d$ shell, and free electrons fill the
band states up to the Fermi energy. 
This ground state is sixfold
degenerate due to the total spin $S=5/2$ of the $3d$ shell. 
In the excited states of the system one electron is
transferred from $3d$ shell to an unoccupied band state, or one
electron from the band is transferred to the $3d$ shell. Such transfers require
excitation energy denoted by $\epsilon_0$, 
which will be the only fitting parameter for 
comparison of MAE given by the effective 
Hamiltonian with the DFT results.

In the excited states there are four or six electrons on the $3d$ shell, 
and according to the Hund's rule their total angular momentum $L=2$ 
and the total spin $S=2$. 
Due to nonzero total angular momentum, in the excited states the internal
Mn spin-orbit coupling $\sim \mathbf{L}\cdot \mathbf{S}$ should be
taken into account. The inclusion of the internal S-O coupling is crucial 
for the splitting of the ground state and magnetic anisotropy properties of the 
ion, otherwise the ground state is not split, independently whether 
the band spin-orbit coupling is present or not. 

In the Hamiltonian of the system the ground and excited states are connected by
hybridization between $3d$(Mn)  orbitals and orbitals of neighbouring 
Te ions.
The effective spin Hamiltonian for the ground state is calculated within the
framework of the second order perturbation theory for degenerate spectra with
respect to hybridization. 

The main differences between the present calculations and those presented in
Ref.~\cite{lusakowski2} for \pmt\ are related to the tight
binding model of the band structure and to the hybridization parameters between
$3d$ states of Mn ion and the orbitals of neighboring Te ions used in 
calculations. In
Ref.~\cite{lusakowski2} we considered Mn ion in PbTe using 
the tight binding model that takes into account nearest cation - anion 
neighbors and 
nearest and next nearest cation - cation and anion - anion neighbors. The
model was parametrized according to Ref.~\cite{kriechbaum}. The
integrals 
describing  hybridization Mn $3d$ -Te orbitals  were taken from
Ref.~\cite{masek}. Here,  the tight binding model takes into
account integrals up to the third neighbour, and their values were taken from 
DFT
calculations for GeTe as the matrix elements of Kohn -- Sham Hamiltonian 
between pseudoatomic orbitals of Ge and Te. 
Using {\it ab initio} OpenMX package \cite{openmx} we performed 
calculations for both fully relativistic  and scalar relativistic 
pseudopotentials. Thus, we can compare effective spin Hamiltonian 
obtained for situations with and without the band spin-orbit interaction. 

From DFT calculations performed for $2\times 2\times
2$ supercell containing 64 atoms with one Ge  replaced by Mn atom we obtain
hybridization parameters between  $3d$(Mn) states and 
$s$ and $p$ orbitals of the nearest Te ions. 

In calculations, like in the paper I,  the crystal structure parameters were 
assumed $a_0=5.98$~\AA,  $\alpha
= 88^{o}$ and $\tau = 0.03$.

Having the effective spin Hamiltonian we obtain 
$E_A(\theta, \varphi)$, i.~e., the dependence of single ion anisotropy energy 
on the 
direction of Mn spin defined by the angles $\theta$ and $\varphi$. 
To this aim we introduce a matrix $D(\theta, \varphi)$
\begin{equation}
 D(\theta, \varphi) = D_z(\varphi)D_y(\theta), 
 \label{eq4}
\end{equation}
where $D_z$ and $D_y$ are rotation matrices for spin $S=5/2$ around the $z$ and 
the $y$ 
axis, respectively. Now, we denote by $|a>$ a state of the Mn spin for which 
the projection on the quantization axis $z$ is $S_z=5/2$, i.e., 
in this state the Mn spin points in the $z$ direction. Then  $|a'> = D(\theta, 
\varphi)|a>$ denotes the state in which the spin points in the direction $z'$ 
defined by the angles $\theta$ and $\varphi$. We define
\begin{equation}
 E_A(\theta, \varphi) = <a'|H|a'>
 \label{eq5}
\end{equation}
as the angular dependence of the single ion anisotropy energy on the spin's 
direction. 
For example, for the Mn ion in perfect cubic environment the effective spin 
Hamiltonian, Eq.~(\ref{eq2}), leads to 

\begin{equation}
E_A(\theta, \varphi) = \frac{5}{4}a\left[(\cos\theta)^4+(\sin 
\theta)^4\left((\cos\varphi)^4 + (\sin\varphi)^4\right)\right]
 \label{eq6}
\end{equation}
which is identical to that obtained by Yosida and Tachiki  
using a different approach \cite{yosida}.

\subsection{Determination of transfer energy $\epsilon_0$}
\begin{figure}
\includegraphics[width=\linewidth]{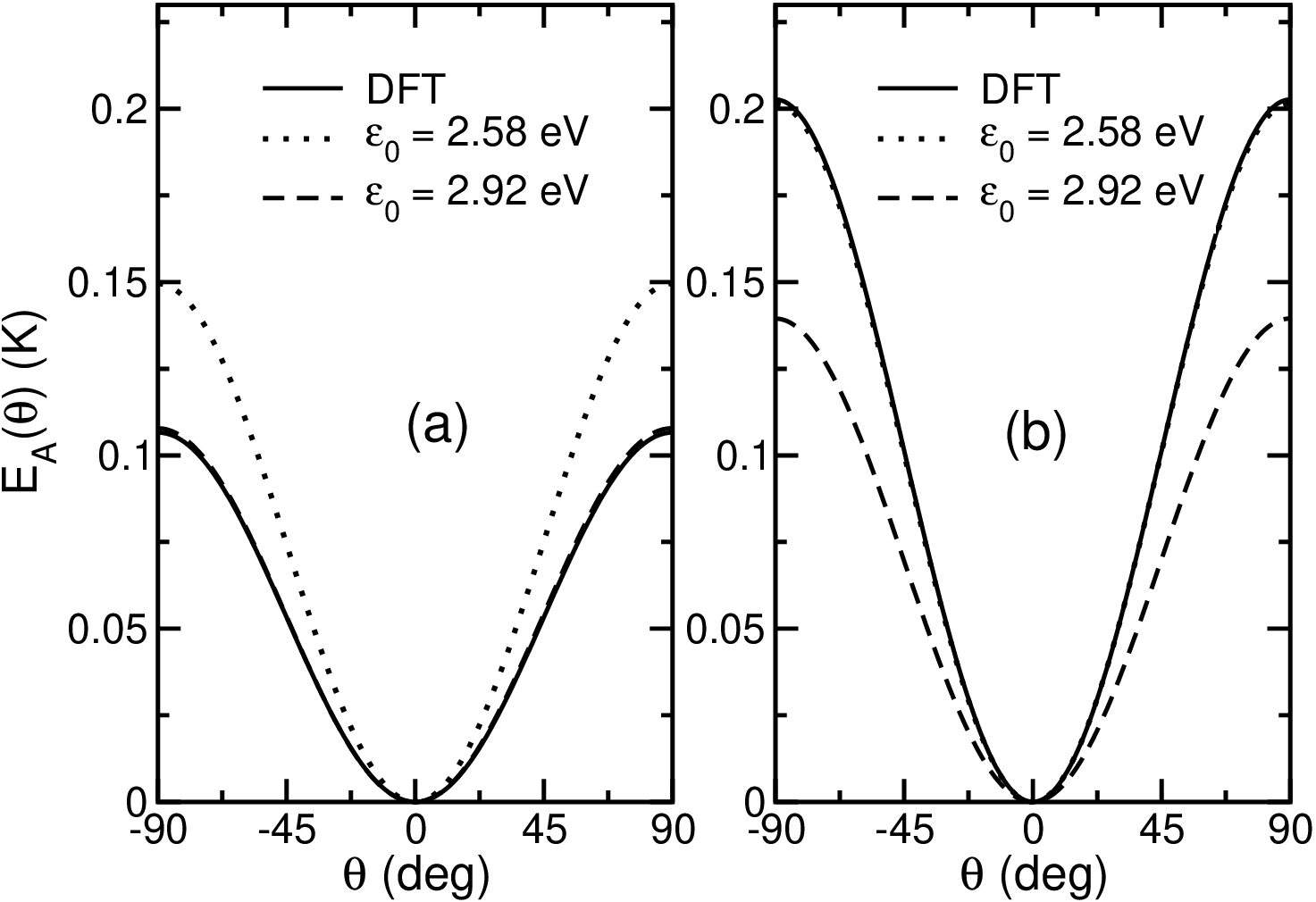}
\caption{%
Angular dependence of MAE for $p=0$ (a) and $p=10^{21}$~cm$^{-3}$ (b).
Continuous lines correspond to DFT calculations. Dotted and broken lines are
single ion anisotropies for two values of transfer energy $\epsilon_0$. }
\label{fig1}
\end{figure}

To determine the energy $\epsilon_0$, i.~e. the energy necessary to
transfer an electron between the $3d$ shell and the top of the valence band, 
we performed scalar 
relativistic calculations for GeTe. Next, using the tight binding 
parameters from the obtained Kohn - Sham Hamiltonian we calculated the 
effective spin 
Hamiltonian and the angular dependence of MAE, $E_A(\theta, \varphi)$, for 
for hole concentrations $p=0$ and $p=10^{21}$~cm$^{-3}$ for different values of 
$\epsilon_0$. The results were compared to MAE 
obtained from DFT calculations for 64-atom supercell containing one Mn ion for
vanishing S-O coupling for Ge and Te atoms. 
Such a comparison makes sense only when the
band spin-orbit interaction is neglected because, as we know from paper I, only
in such a case MAE is due to Mn ion and does not depend on the direction
of spin polarization of Ge and Te ions. 
From DFT calculations we obtain $a_{2c}=-0.053$~K and $a_{2c}=-0.101$~K for 
insulating and 
conducting cases, respectively. To obtain these values by the 
effective Hamiltonian method we must assume $\epsilon_0$=2.92~eV for the 
insulating, and 
$\epsilon_0$=2.58~eV for the conducting cases, respectively, see the Fig.
\ref{fig1}. Because in the 
following we consider the conducting case, we assume $\epsilon_0$=2.58~eV. The 
value
of $\epsilon_0$ obtained from fitting of the present model to DFT {\it ab
initio} calculations are quite reasonable, and comparable with the values used
previously in the literature for PbTe.
In Ref.~\cite{lusakowski2} $\epsilon_0$ was considered in the interval
from 1.6~eV to 3.5~eV, in Ref.~\cite{sliwa} value 3.5~eV was assumed.

\subsection{Dependence of magnetization easy axis direction on disorder}
\begin{figure}
\includegraphics[width=\linewidth]{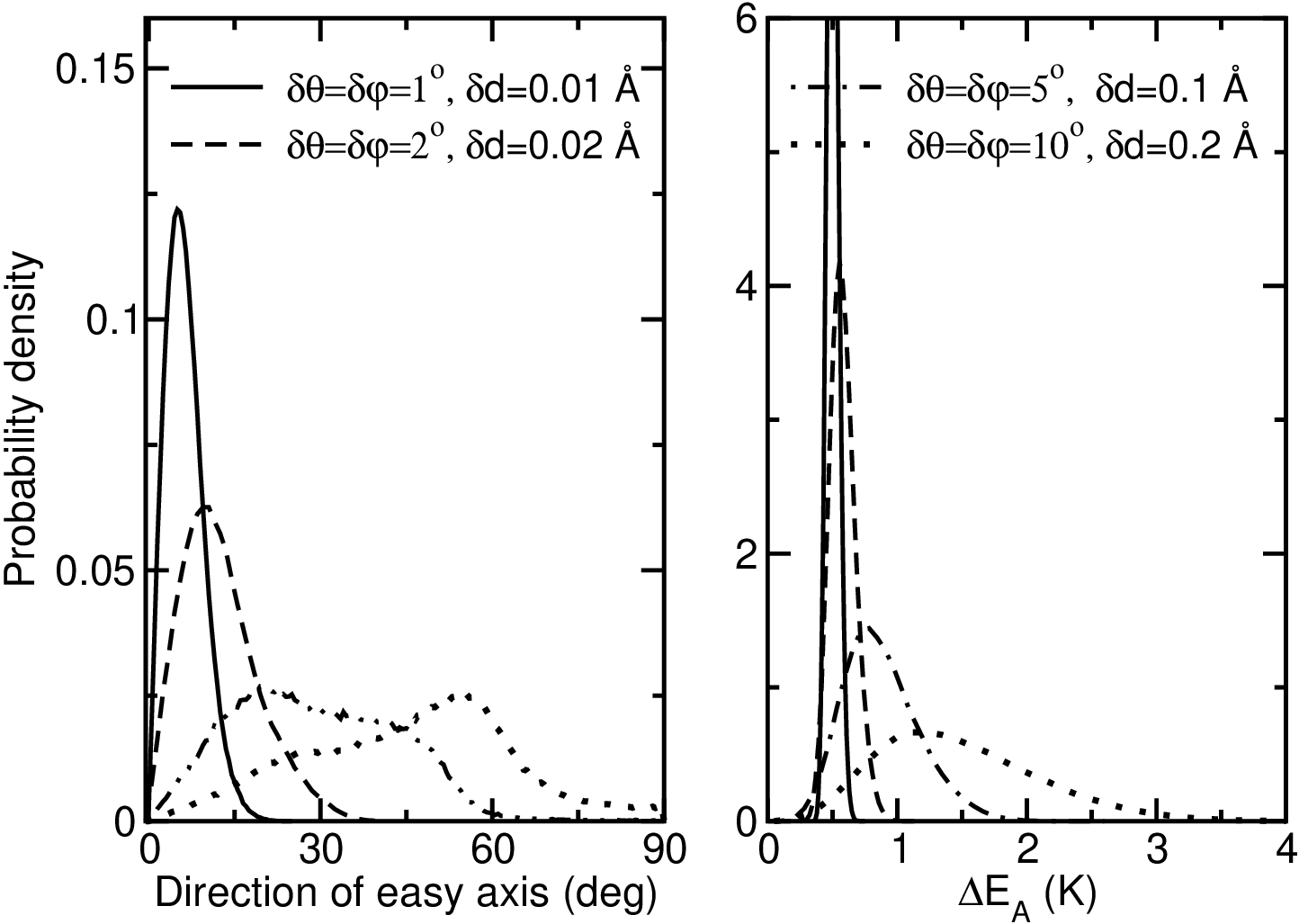}
\caption{%
Probability density for the angle between easy axis and [111] direction (a) and
for anisotropy energies (b) for different levels of disorder quantified by the 
spreads of the angles $\delta \theta$, $\delta \varphi$ and bond length $\delta 
d$.}
\label{fig2}
\end{figure}
To simulate microscopic disorder in \gmt\  we start from the 
perfect GeTe lattice characterized by $a_0=5.98$~\AA, $\alpha=88^o$, 
$\tau=0.03$ with one Ge replaced by Mn. The  
directions and lengths of six bonds connecting  Mn with its nearest Te
neighbours are specified by $(\theta^0_i, \varphi^0_i, d^0_i)$, 
$i=1,...,6$. 
We assume that the microscopic disorder leads to small changes in $(\theta_i, 
\varphi_i, d_i)$,
\begin{eqnarray}
 \theta_i = \theta^0_i+r_i^{\theta}\delta\theta \nonumber\\
 \varphi_i = \varphi^0_i+r_i^{\varphi}\delta\varphi\\
 d_i = d^0_i+r_i^{d}\delta d \nonumber
 \label{eq7}
\end{eqnarray}
where $r_i^{\theta}$, $r_i^{\varphi}$, $r_i^{d}$ are homogeneously distributed  
random numbers from the interval (-0.5, 0.5). In calculations it is assumed
that Mn-Te hybridization parameters change with the bond's
length as $d^{-7/2}$ \cite{harrison}.

For a given configuration of Mn surrounding we calculate effective spin 
Hamiltonian, and using Eq.~(\ref{eq5}) we find $E_A(\theta, \varphi)$. 
Having this functional dependence we obtain direction of the easy axis of 
magnetization and $\Delta E_A$, the difference between maximal 
and minimal values of $E_A(\theta, \varphi)$. In Fig.~\ref{fig2} we present 
normalized to unity histograms (probability densities) of directions of the 
easy 
axis of magnetization with respect to [111] crystallographic direction, 
Fig.~\ref{fig2}a, and of
$\Delta E_A$, Fig.~\ref{fig2}b. These histograms  result from $4\times 10^5$ 
random configurations.

To obtain typical deflections of directions and lengths of Mn-Te
bonds from those for the ideal lattice we performed, using OpenMX
package \cite{openmx}, geometry optimization for
the $2\times 2\times 2$ 64-atom supercell. 
 We repeated the
optimization for three different cases:  1) with one Mn ion, 2) with two Mn ions
placed at
nearest neighbours positions and 3) with two Mn ions placed at the two diagonal
vertices of the elementary cell and with one Ge vacancy on the face of the
elementary cell.

Comparing the directions and lengths of Mn-Te bonds between cases 2) and 3)
with those in 1) we find that the the deflections are very small. The
highest deflections are of the order of $1^o$ and changes
of bond lengths are of the order of 0.01~\AA. 
We see that even for very small deflections of directions and lengths of bonds
from the ideal values, the influence on the distribution of the directions of 
the easy 
axes is quite substantial. However, the probability distribution for $\Delta
E_A$ for such small $\delta\theta$, $\delta \varphi$ and $\delta d$ remains
well centered around the mean value, which is of the order of 0.5~K with the 
variance less than 0.1~K. 

We have also calculated the averaged over disorder the
magnetic field dependent magnetizations of Mn ion with magnetic field along 
[111] and perpendicular to [111]
direction in different temperatures. The differences are noticeable, however
in very small magnetic fields and extremely low temperatures only. The
differences which might be compared to those experimentally observed are only
for unphysically high level of disorder, see Fig.~\ref{fig2}. 

We conclude therefore that the single ion anisotropy does not play a decisive 
role in
magnetic anisotropy energy in \gmt. The present calculations further confirm the
conclusion of the paper I that MAE in \gmt\ is mainly due to polarization of
valence band carriers.

\section{Supercells containing more than one M\lowercase{n} ion}
\label{sec3}
In the present Section we analyze magnetic anisotropy for all possible  
configurations of two Mn ions in the $2\times 2\times2$ supercell, 
for the case of three Mn ions we show only results for a few randomly chosen 
configurations. 
Technical details of calculations were presented in the paper I. 
We used OpenMX package \cite{openmx}. In 
all calculations of MAE the number of integration points in the Brillouin zone 
were 512 and 2744 for insulating and conducting cases, respectively. As was 
shown in the paper I such numbers give convergent results. 
In the conducting case, the calculations were performed for hole concentration 
$p=10^{21}$~cm$^{-3}$, which is typical for samples studied experimentally. 
For geometry optimization, the number of integration points in the 
Brillouin zone reduced to 64 give convergent results. The force criterion for 
geometry 
optimization was equal to $5\times 10^{-3}$ Hartree/Bohr. 

In order to describe possible configurations, first, we introduce notation
describing positions of Mn ions in the supercell.\\ 
Let us
consider three vectors ${\bf c_1}$, ${\bf c_2}$ and ${\bf
c_3}$, parallel to the [100], [010] and [001] crystallographic directions,
respectively. The lengths of the vectors are the same, $a_0/2$. 
The symbol \{{\em ijk}\} describes the position of an atom in the supercell 
${\bf
r}=i{\bf c_1}+j{\bf c_2}+k{\bf c_3}$ where $i$, $j$, $k$ are integer numbers. 

\begin{figure}
\includegraphics[width=\linewidth]{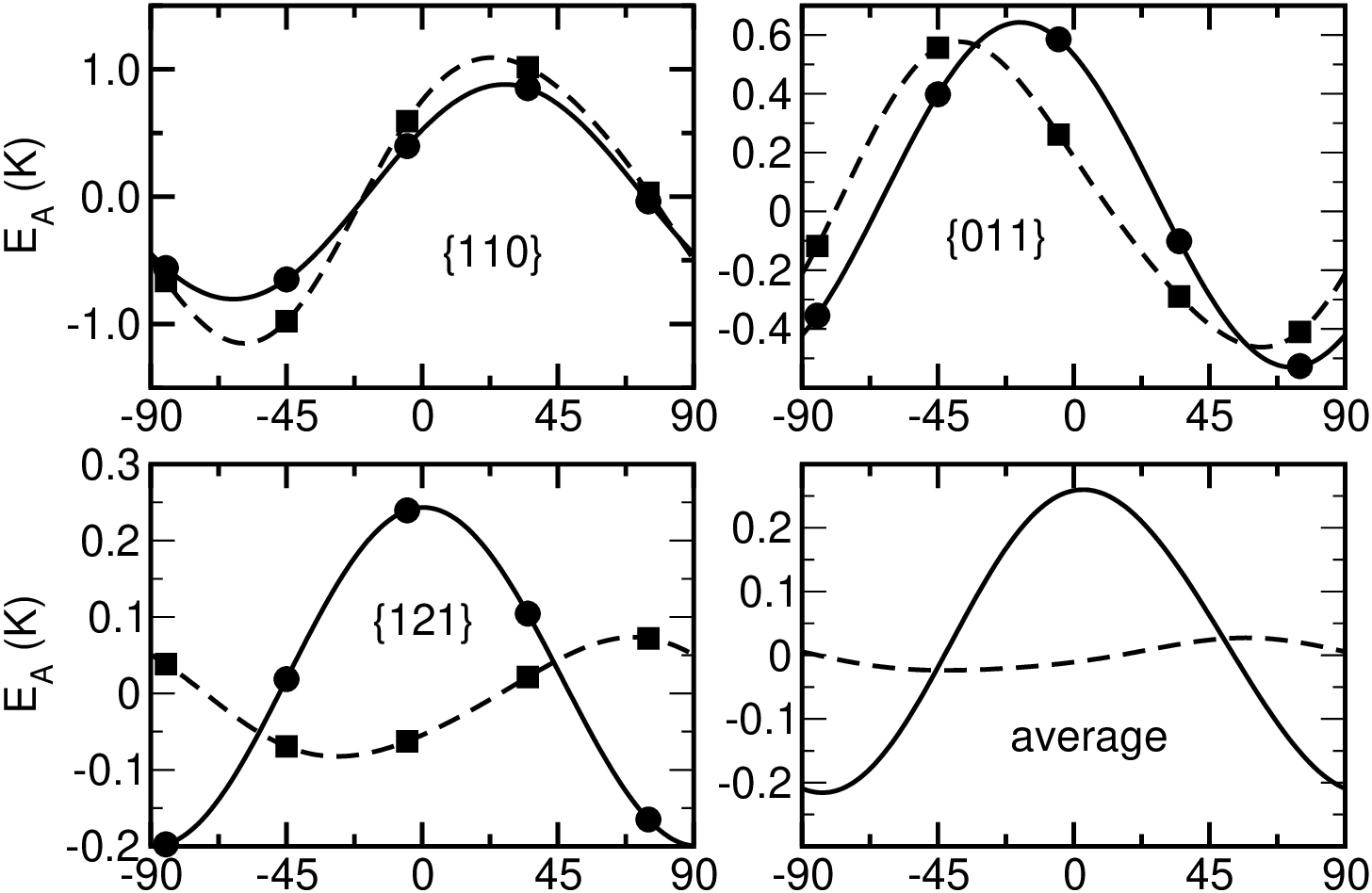}
\caption{%
Angular dependence of the energy of magnetic anisotropy for $2\times 2\times 2$
supercell containing two Mn ions for insulating case. Continuous lines -
nonrelaxed lattice, broken lines - relaxed lattice. 
}
\label{fig3}
\end{figure}

The first Mn atom is always placed at \{000\}. 
The $2\times2\times2$ supercell contain 32 different cation positions, thus the
second Mn atom may be placed at 31 remaining positions. However, some of the
resulting 31 configurations are equivalent because of the periodic boundary
conditions; for example configurations \{110\} and \{330\} are equivalent. 
Actually, there are 21 nonequivalent configurations. 
We stress that due
to nonzero Mn spin this number is larger then the number resulting from
purely geometrical symmetry considerations for a rhombohedral lattice. Consider,
for example, three nearest neighbor pair 
configurations for which the second Mn atom is placed at \{110\}, \{011\} and
\{101\}, respectively.  For vanishing spin, those three configurations are
equivalent from the purely geometrical point of view, being related by the
rotation along [111] axis by the angle $\pm 120^{\circ}$. However, for finite 
$S$, 
they are nonequivalent if we consider magnetic properties of the crystal, what
is explicitly shown in the following in Fig.~\ref{fig3}. 

In the calculations,  the Mn spin vectors are placed in the (-110)
plane. Due to reflection symmetry with respect to this plane,
the configurations \{011\} and
\{101\} give identical results for the angular dependence of
$E(\theta)$, thus it is enough to perform calculations
for one of them, but for arbitrary direction of manganese spins it is 
necessary to calculate MAE for both.

\begin{figure}
\includegraphics[width=\linewidth]{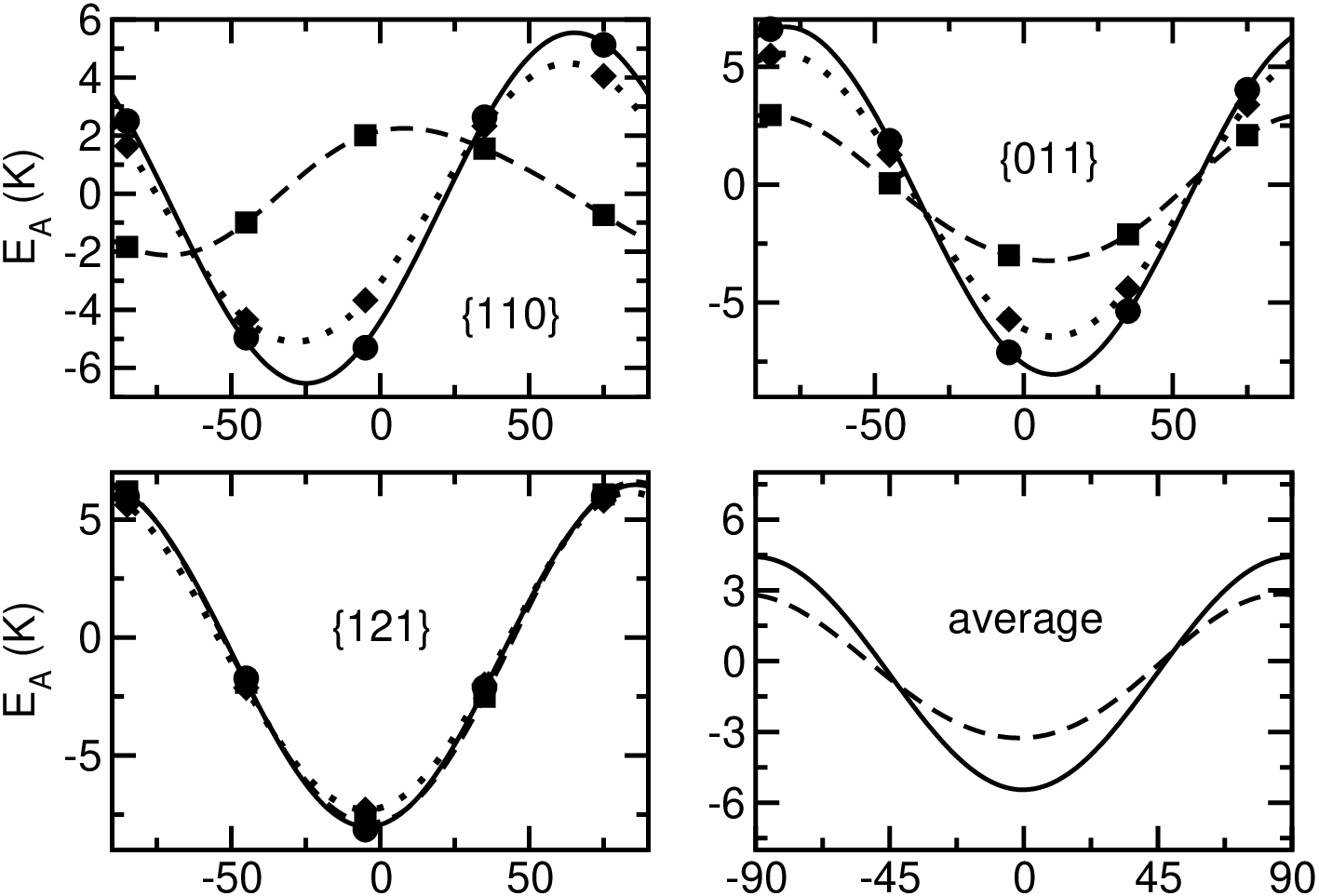}
\caption{%
Angular dependence of the energy of magnetic anisotropy for $2\times 2\times 2$
supercell containing two Mn ions for hole concentration $p=10^{21}$~cm$^{-3}$.
Continuous lines -
nonrelaxed lattice, broken lines - relaxed lattice characterized by lattice
parameters $a_0=5.98$~\AA, $\alpha=88^o$, $\tau=0.03$. The dotted lines are
for nonrelaxed lattice for other lattice parameters $a_0=5.98$~\AA,
$\alpha=88.3^o$, $\tau=0.025$. 
}
\label{fig4}
\end{figure}
\ \\ \ \\
\begin{figure}
\includegraphics[width=\linewidth]{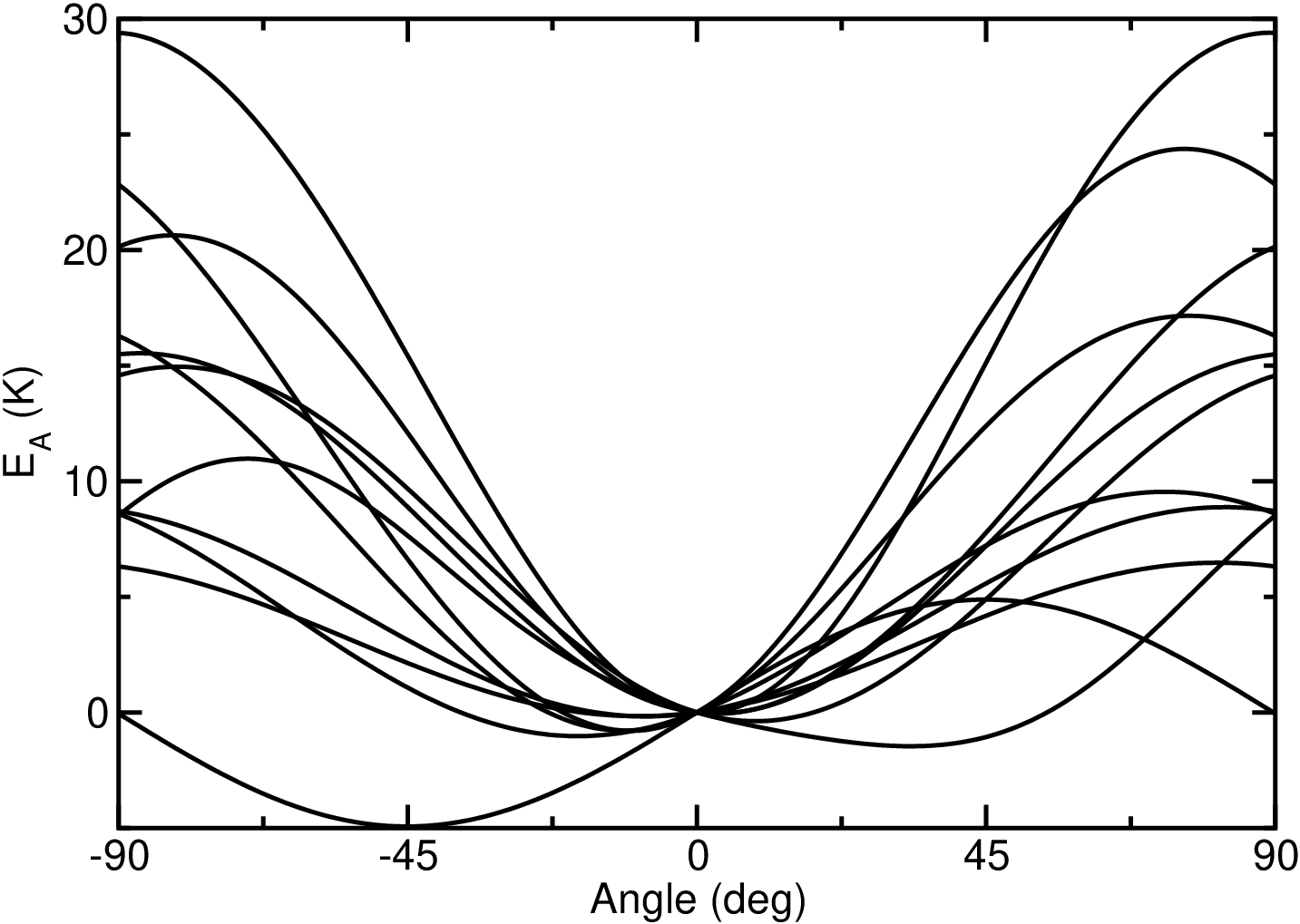}
\caption{%
Angular dependence of MAE for several configurations of three Mn ions in 
$2\times 2\times 2$ supercell for hole concentration
$p=10^{21}$~cm$^{-3}$.
}
\label{fig5}
\end{figure}

\begin{table}
\caption{\label{tab1} Values of coefficients $a_{2c}$ and $a_{2s}$ describing
MAE for $2\times 2\times 2$ supercell containing two Mn ions for  nonrelaxed and
relaxed lattices and hole concentration $p=0$.}
 \begin{tabular}{|c|c|r|r|r|r|}
\hline
  &&\multicolumn{2}{c|}{nonrelaxed}&\multicolumn{2}{c|}{relaxed}\\
  \hline
configuration&$f$&$a_{2c}$&$a_{2s}$&$a_{2c}$&$a_{2s}$\\
\hline
\{110\}&1&0.49&0.69& 0.62 & 0.92\\        
\{011\}&2&0.48&-0.34& 0.20 & -0.47\\        
\{002\}&1&-0.12&-1.13&-0.84 & -1.56\\        
\{112\}&1&0.24&-0.01& -0.11 & -0.10\\        
\{013\}&2&-0.33&-0.42& -0.72 & -0.36\\        
\{020\}&2&0.31&0.57& 0.42 & 0.80\\        
\{130\}&2&0.04&0.96& 0.05 & 0.71\\        
\{121\}&2&0.22&0.01& -0.05 & 0.06\\        
\{022\}&2&0.32&-0.16&-0.10 & -0.27\\        
\{132\}&2&0.36&-0.22&-0.05 & -0.21\\        
\{123\}&2&0.38&0.11& 0.15 & 0.11\\        
\{220\}&1&0.26&0.29& 0.15 & 0.50\\        
\{222\}&1&0.48&-0.01& 0.20 & 0.00\\        
\hline
average&&0.23&0.04&-0.01 & 0.02\\
\hline
 \end{tabular}
\end{table}

\begin{table}
\caption{\label{tab2} Values of coefficients $a_{2c}$ and $a_{2s}$ describing
MAE for
$2\times 2\times 2$ supercell containing two Mn ions for  nonrelaxed and
relaxed lattices and hole concentration $p=10^{21}$cm$^{-3}$.}
 \begin{tabular}{|c|c|r|r||r|r|}
\hline
  &&\multicolumn{2}{c||}{nonrelaxed}&\multicolumn{2}{c|}{relaxed}\\
  \hline
configuration&$f$&$a_{2c}$&$a_{2s}$&$a_{2c}$&$a_{2s}$\\
\{110\}&1&-3.91 & 4.6& 1.88 & 1.03\\        
\{011\}&2&-6.93 & -2.51& -3.02 & -0.61\\        
\{002\}&1&-3.81 & 0.68& -1.55 & 2.14\\        
\{112\}&1&-7.31 & -1.92& -6.39 & -1.12\\        
\{013\}&2&-3.02 & -1.84& -4.22 & -2.96\\        
\{020\}&2&-3.13 & -0.46& 1.27 & -1.02\\        
\{130\}&2&-0.29 & 3.74& -0.42 & 5.92\\        
\{121\}&2&-7.16 & 1.09& -7.16 & 0.98\\        
\{022\}&2&-7.47 & -3.22& -6.56 & -3.96\\        
\{132\}&2&-6.51 & -2.35& -4.08 & -2.16\\        
\{123\}&2&-5.67 & 1.06& -2.42 & 1.53\\        
\{220\}&1&-4.87 & 6.39& -2.84 & 8.11\\        
\{222\}&1&-3.36 & 0.11& -1.48 & 0.48\\        
\hline
average&&-4.93 & 0.04& -3.03 & 0.29\\
\hline
 \end{tabular}
\end{table}

In Figs. \ref{fig3} and \ref{fig4} we present the results for the 
chosen example configurations for insulating and conducting cases,
respectively. Also we show arithmetical averages over all
nonequivalent configurations.  Let us notice, however, that the results 
presented in the paper I indicate that 
even in the insulating case the pair of Mn ions in the $2\times
2\times 2$ supercell cannot be
treated as isolated.  Thus, strictly speaking, these arithmetical averages
cannot be treated as the 
averages over the quenched disorder. Nevertheless,  they give some information 
about
behavior of MAE in macroscopic crystal where the magnetic ions are placed
randomly. 

In Figs. \ref{fig3} and \ref{fig4} we compare also MAE for nonrelaxed
(continuous lines) and relaxed (broken lines) supercells. As in the case of
single Mn ion in the supercell, the relaxation of the lattice leads to
significant changes in MAE. 
For comparison, by the dotted lines we show MAE for
nonrelaxed lattice with slightly different lattice parameters:  $a_0=5.98$~\AA,
$\alpha=88.3^o$, $\tau=0.025$. Like in paper I, for single Mn ion in the
supercell, such changes result in changes of MAE.

The results for all configurations are summarized in
Tables \ref{tab1} and \ref{tab2}.
The first column of each table gives configurations, in the second column 
there are corresponding
weight factors $f$. The factor $f=1$ is for a configuration which remains
the same when reflected in the (-110) plane,  and $f=2$ if there exist two 
nonequivalent configurations related  by reflection in the (-110)
plane. The coefficients $a_{2c}$ and $a_{2s}$ describe curves used to fitting
the calculated points:
\begin{equation}
 E_A(\theta) = a_{2c}\cos(2\theta) + a_{2s}\sin(2\theta)
 \label{eq3}
\end{equation}
It turns out that inclusion of terms proportional to
$\cos(4\theta)$ and $\sin (4\theta)$ to the fitting formula 
does not improve  quality of the fit, because the
coefficients describing these terms are more than order of magnitude
smaller than $a_{2c}$ and $a_{2s}$. 

%

From the presented results we draw two main conclusions.\\
First, for a given pair configuration the direction of the easy axis of
magnetization, in general, is very different from [111] crystallographic
direction or the direction perpendicular to [111]. However, after the assumed
averaging procedure the average of the $a_{2s}$ coefficient is very small, it
means that the average MAE possesses rhombohedral symmetry of the macroscopic
crystal. Similar behavior was previously observed in calculations of single ion
magnetic anisotropy in \smt\ \cite{lusak_ssc} in which disorder effects were
taken into account.\\
Second, in most of cases the values of $\sqrt{a_{2c}^2+a_{2s}^2}$ for pair
configurations are significantly larger than the average of coefficient
$a_{2c}$. This fact is also connected with varying directions 
of the
easy axes of magnetization for different pair configurations. 

In Fig. \ref{fig5} we show results for several randomly chosen configurations
of
three Mn ions in $2\times 2\times 2$ supercell. We are not able to perform
calculations for all possible configurations because in this case the number of
configurations is too large, however one may notice that the qualitative
results are similar to the case of two Mn ions in the supercell. 

\begin{figure}
\includegraphics[width=\linewidth]{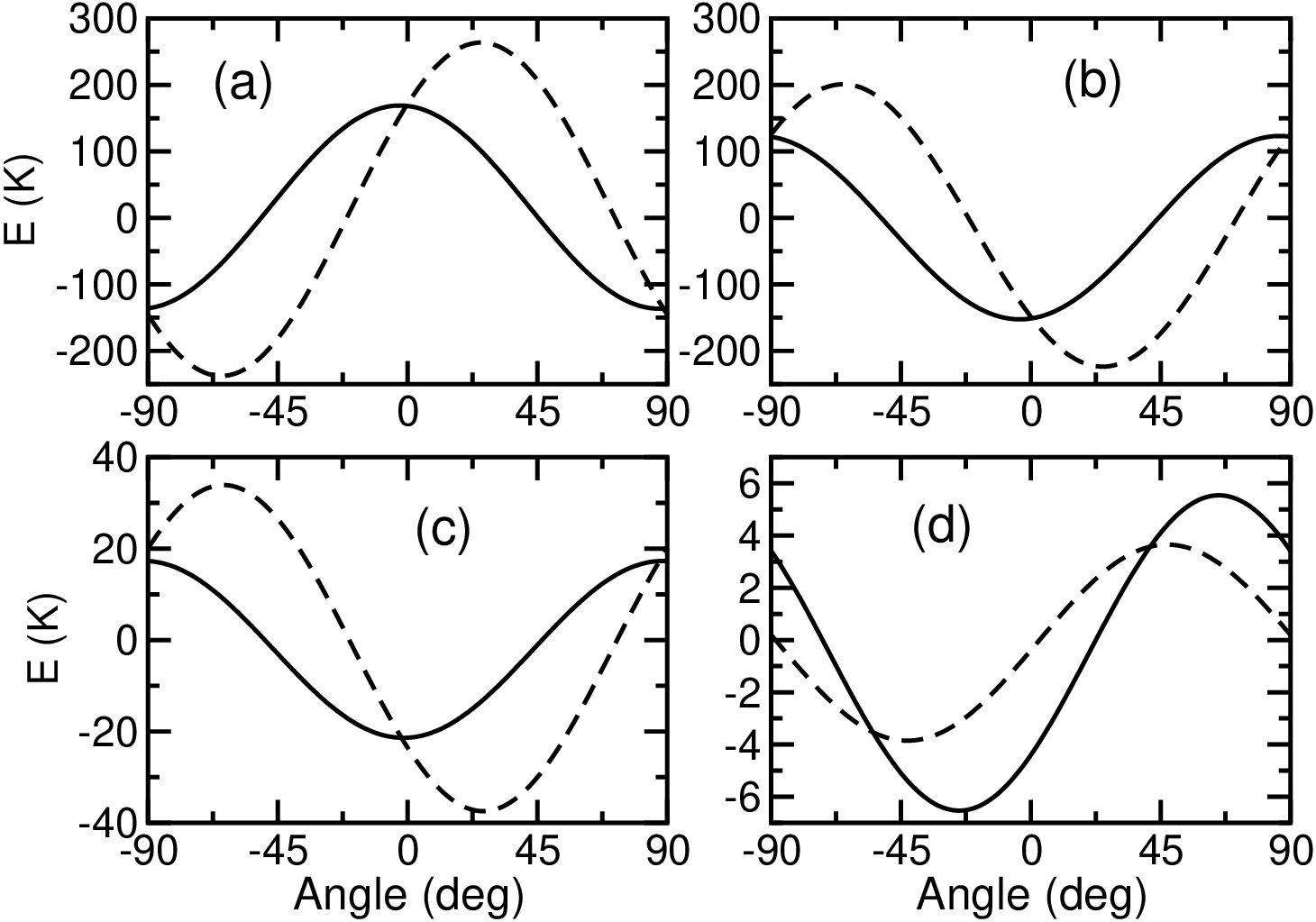}
\caption{%
Angular dependence of contributions of kinetic (a), electrostatic (b) and
exchange-correlation (c) energies to magnetic anisotropy energy (d) for
$2\times 2\times 2$
supercell containing two Mn ions in \{110\} (continuous lines) and \{130\}
(broken lines) configurations. The hole concentration is $p=10^{21}$~cm$^{-3}$.
\label{fig6}
}
\end{figure}

Angular dependencies for the kinetic, electrostatic and
exchange-correlation components of MAE are shown in Fig.~\ref{fig6}. 
Their nonconstant angular dependencies prove that the spatial 
charge distribution varies with the Mn spin's direction.

Two features may be noticed. First, the changes of kinetic, electrostatic and
exchange-correlation energies with the direction of manganese spins are much
larger than the changes of their sum, the magnetic anisotropy energy. However 
this is usual in DFT calculations - the
dependence of components of the total energy on external parameters is much
stronger than the dependence of the total energy itself. The other feature is 
more
interesting. Let us notice that the maxima and minima of components are placed
in different positions than the maxima and minima of MAE. This was not the
case for single manganese ion in a supercell considered in the paper I where the
maxima or minima for all curves were at the same positions. This shows, 
even more expressively than in the paper I, the close relationship between spin 
direction of manganese ions and the spatial distribution of electron charge. 
The changes of spatial charge density are caused by the spin-orbit 
interaction.

\section{Conclusions}
\label{sec4}
In this work, we theoretically analyzed the influence of chemical disorder on 
magnetic anisotropy in ferromagnetic \gmt\  semiconductor layers known to 
undergo ferroelectric structural distortion of rhombohedral crystal lattice and 
exhibit perpendicular magnetic anisotropy. Using DFT method we determined the 
local crystal structure in \gmt\  substitutional solid solution and 
calculated the single ion contribution to magnetic anisotropy energy. Our 
calculation revealed that the single ion contribution is order of magnitude 
smaller than the contribution originating from spin polarization induced by Mn 
ions into  neighboring Te and Ge ions. This effect results from the 
hybridization of magnetic $3d$ orbitals of Mn ions and valence band states of 
GeTe subject to spin-orbit interactions \cite{lusakowski0}.

We also discussed magnetic
anisotropy effects due to pairs of Mn ions differently allocated in the 
supercell and showed that the spatial averaging over the chemical disorder 
strongly
reduces the strength of this magnetic anisotropy mechanism and restores the
global rhombohedral symmetry of magnetic system.

Finally, the calculations show also that although in principle {\it
ab initio} numerical methods may be applied to the calculations of MAE, in order
to take into account disorder and make a direct quantitative comparison with
experiment, much larger systems must be considered.

\section*{Acknowledgement} 
The authors acknowledge the support from NCN (Poland) research project 
no. UMO-2011/01/B/ST3/02486. This research was supported in part by PL-Grid
Infrastructure. 

\end{document}